\documentclass[11pt,letterpaper]{article} 
\usepackage{ol2}
\usepackage[draft]{hyperref} 
\usepackage{srcltx}

\usepackage{srcltx}

\usepackage[draft]{hyperref} 

\begin{document}

\title{Holographic microscopy for the three-dimensional exploration of light scattering from gold nanomarkers in biological media}

\author{Fadwa Joud$^a$, Fr\'{e}d\'{e}ric Verpillat$^a$, Pierre Desbiolles$^a$, Marie Abboud$^c$ and Michel Gross$^b$}

\affiliation{
$^a$ Laboratoire Kastler Brossel, Ecole Normale Sup\'{e}rieure, UPMC, 24 Rue Lhomond, 75005
Paris, France;\\
$^b$ Laboratoire Charles Coulomb, Universit\'{e} Montpellier II, Place Eug\`{e}ne Bataillon, 34095
Montpellier, France;\\
$^c$ D\'{e}partement de Physique, Facult\'{e} des Sciences, Universit\'{e} Saint-Joseph, Beirut, Lebanon
}

\begin{abstract}
The 3D structure of light scattering from dark-field illuminated live 3T3 cells marked with 40 nm gold nanomarkers
is explored. For this purpose, we use a high resolution holographic microscope combining the off-axis heterodyne
geometry and the phase-shifting acquisition of the digital holograms. Images are obtained using a novel
3D reconstruction method providing longitudinally undistorted 3D images. A comparative study of the 3D
reconstructions
of the scattered fields allows us to locate the gold markers which yield, contrarily to the cellular
structures, well defined bright scattering patterns that are not angularly titled and clearly located along the optical
axis. This characterization is an unambiguous signature of the presence of the gold biological nanomarkers,
and validates the capability of digital holographic microscopy to discriminate them from background signals in
live cells.
\end{abstract}

\maketitle

\bigskip

\section{Introduction}

Digital Holographic Microscopy is gaining increasing interest in biological research \cite{charriere2006cell,mann2006movies}   because it enables the 3D
reconstruction of both the amplitude and the quantitative phase images from a single recorded hologram and
without any mechanical scanning. Since Gold nanoparticles (NPs) are non cytotoxic and benefits from interesting
optical properties, their use as biological markers is advantageous \cite{el2005surface,lasne2006single}. In this communication, we used a new
holographic microscope that combines the off-axis geometry\cite{leith1965microscopy} and phase-shifting acquisition of holograms \cite{zhang1998three} based
on an original 3D reconstruction-without-distortion method to image live 3T3 mouse fibroblasts marked with 40
nm gold NPs. The proposed reconstruction method is a two step method providing longitudinally undistorted
3D reconstructed images and corrected for the off-axis tilt and lens aberrations.
We showed that important information can be derived not only from the intensity of the bright spots caused
by the gold NPs, but also from the 3D shape of the light scattering pattern, which is easily accessed using
holography. We showed, in particular, that the speckle signal keeps memory of the illumination direction, while
the particle signal does not.

\section{Methodology}

\subsection{Sample preparation}

\begin{figure}[h]
\begin{center}
  \includegraphics[width=8 cm]{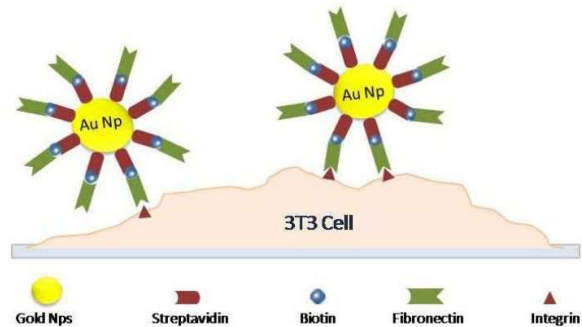}\\
  \caption{Biological specimen preparation.}\label{Fig1}
\end{center}
\end{figure}

We use cultures of NIH 3T3 mouse fibroblasts incubated with 40 nm gold bioconjugates. Gold NPs were
previously functionalized with Fibronectin proteins allowing their bonding with the Integrin surface cellular
receptors of the cells (See Fig. \ref{Fig1}). The bio-conjugation and functionalization protocols that we used to prepare
the biological samples are the same protocols described in details in Ref. \cite{warnasooriya2010imaging}.
Novel Biophotonic Techniques and Applications, edited by Henricus J. C. M. Sterenborg, I. Alex Vitkin, Proc. of

\subsection{Optical setup}

\begin{figure}[h]
\begin{center}
 \includegraphics[width=8 cm]{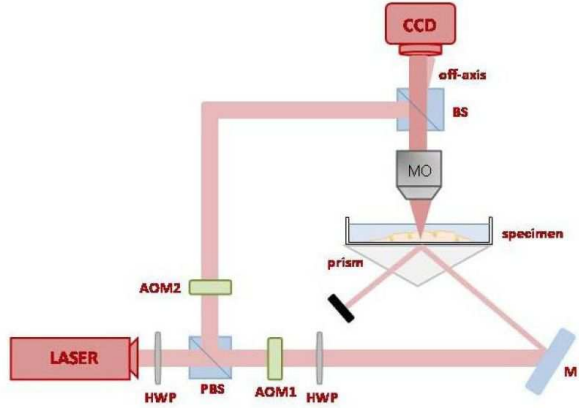}\\
  \caption{Optical setup, M: Mirror, HWP: Half Wave Plate.}\label{Fig2}
\end{center}
\end{figure}

The optical setup is a modified Mach-Zender interferometer (Fig. \ref{Fig2}). Coherent light is produced by a laser source
which is divided by a polarizing beam splitter (PBS). One of the two splitted beams is used to illuminate the
sample in Total Internal Reflection (TIR). A microscope objective (MO) collects the scattered light and forms
the object wave which interferes with a reference beam to produce the hologram that is recorded by a digital
CCD camera.

\subsection{Numerical reconstruction}

Holograms are recorded in an off-axis geometry.5 Phase-shifting interferometry is performed using two synchronized
Acousto-optic modulators (AOM), one in each arm of the interferometer, allowing independent frequency
modulation \cite{atlan2007accurate,gross2008noise,atlan2008heterodyne}. The combination of off-axis and phase-shifting is an effective configuration to reduce parasitic
noise and eliminate images aliases.9 The hologram, carrying both amplitude and phase information, is then
numerically treated and reconstructed using our novel reconstruction-without-distortion method that uses two
Fourier Transforms (FFT). A first FFT is performed to obtain a clear bright disk which is the image of the MO
exit pupil plane in the reciprocal space. Since the light scattered by the sample is collected by the MO, all interesting
information is included inside this disk. A circular spatial mask is thus applied to eliminate all parasitic
signal and keep only this interesting signal. The off-axis tilt is then compensated by translating the circle to the
center of the reciprocal space. Using two FFT \cite{le2000numerical}, the object field at the object plane is then reconstructed from
the image, through the MO, of the recorded hologram at the CCD plane. 3D images are obtained by performing
this 2FFT reconstruction at varying reconstruction distances. Because the propagation is done in free-space, the
pixel size is kept constant and the method do not suffer from longitudinal distortions of the 3D images.

\section{Experimental results}

\begin{figure}
\begin{center}
 \includegraphics[width=12 cm]{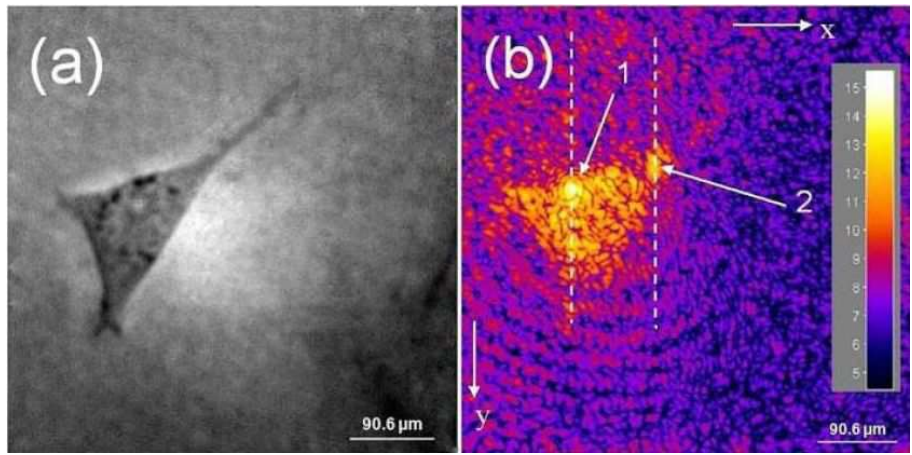}\\
  \caption{Images of 3T3 mouse fibroblast marked with 40 nm gold particle : (a) direct image under white light illumination, (b) reconstructed holographic intensity image in logarithmic scale.}\label{Fig3}
\end{center}
\end{figure}

\begin{figure}
\begin{center}
   \includegraphics[width=12 cm]{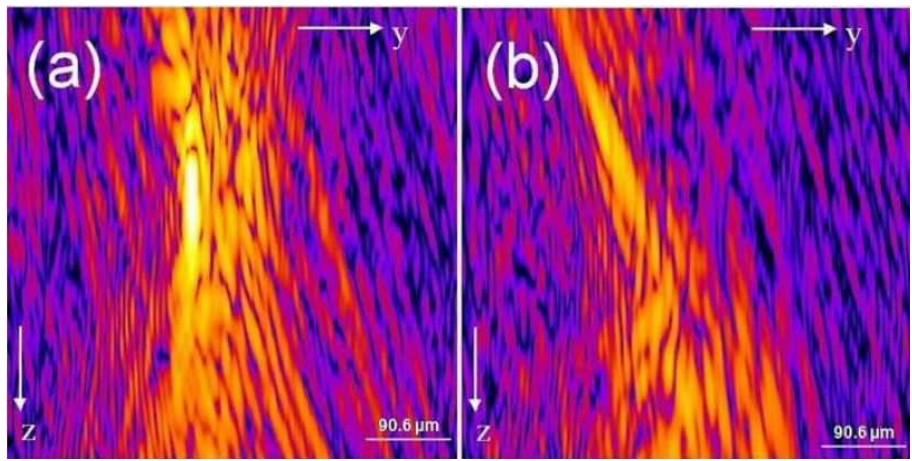}\\
  \caption{ Cuts along the x plane : (a) the cut corresponds to the white dashed line 1 of Fig.2(a), (b) the cut corresponds to the white dashed line 2 of Fig. \ref{Fig2}(a). }\label{Fig4}
\end{center}
\end{figure}

We imaged a 3T3 cell marked with 40 nm gold bioconjugates. The cell can be clearly seen of Fig. 3(a), which
is a direct brightfield image. Fig. 3(b) is the reconstructed holographic image. The brightest point (arrow 1 on
Fig. 3(b)) is a gold marker and the bright point marked by arrow 2 is a speckle hot spot. To better characterize
the particles signal with respect to hot spots, we have analyzed the 3D images of the wave-field obtained by
reconstruction for 512 different reconstruction planes and we performed cuts, at the position of the gold particle
(point 1) and at the position of the hot spot (point2), along the x plane that is parallel to the yz incidence plane
of the sample illumination beam. We can see on the scattering patterns images of Fig.4 that, contrarily to the
particle, the hot spot signal extension along the vertical axis is quite large. Moreover, the hot spot image is
angularly tilted in the $yz$ plane while the particle signal is a well defined straight bright pattern clearly located
along the optical axis z. This effect is a direct consequence of the non-similarity of the angular distribution of
the light scattered from biological tissues and from gold nanoparticles, which depends on the anisotropy factor $g$ (with  $g\simeq 1$ for cells) giving a forward scattering regime while gold nanoparticles scatters light isotropically since
their anisotropy factor is null \cite{cheong1990review}.

\section{Conclusion}

Off-axis holographic microscopy is well adapted to the detection of weakly scattering objects. Image reconstructions
are done using a novel 3D reconstruction-without-distortion method potentially effective for thick volume
objects imagery. The sensitivity, SNR and selectivity of the technique allow the localization of gold nanoparticles
of a few tens of nanometers. Biological environments, however, are difficult to address since cell features generate
strong parasitic speckle. Here, we have reported the detection of 40 nm particles attached to the surface of live
3T3 cells. We show that, in addition to a stronger scattering signal, gold particles induce a relatively isotropic
scattering, whereas biological features are characterized by mostly forward scattering. This dissimilarity in the
scattering patterns, explained by the inconsistency of the refractive indexes and anisotropy parameters g, is
easily characterized by digital holography, making it an excellent tool for the 3D detection of gold markers in
biological environments.

\bigskip

\textbf{Ackowledgments}

Authors wish to acknowledge the French Agence Naionale de la Recherche (ANR) and the Centre de Comp\'{e}tence NanoSciences Ile de France (Cnano IdF) for their financial support.

\bibliographystyle{unsrt}

\end{document}